\documentstyle [fleqn,12pt]{article}
\begin{document}
\baselineskip .75cm 
\begin{titlepage}
\title{ \bf Statistical mechanics of confined quantum particles}      
\author{Vishnu M. Bannur and K. M. Udayanandan\\
{\it Department of Physics}, \\  
{\it University of Calicut, Kerala-673 635, India.} }
\maketitle
\begin{abstract} 

We develop statistical mechanics and thermodynamics of Bose and Fermi 
systems in relativistic harmonic oscillator (RHO) confining potential, which 
may be applicable in quark gluon plasma (QGP), astrophysics, 
Bose-Einstein condensation (BEC)
etc. Detailed study of QGP system is carried out and compared with lattice 
results. Further, as an application, our equation of state (EoS) of QGP  is 
used to study compact stars like quark star.  
 
\end{abstract}
\vspace{1cm}

\end{titlepage}
\section{Introduction :}
  
There are many systems like QGP, quark stars, BEC etc. where particles 
are confined by harmonic oscillator type potentials and are at thermodynamic 
equilibrium. So we need to study statistical mechanics of such a system. 
Taking QGP, a new form of matter made of quarks and gluons, as an example 
we discuss this problem.   
It is generally accepted that hadrons are confined state of quarks and 
gluons and at high temperature and density, we get what is called QGP 
\cite{qgp.1}. 
In relativistic heavy ion collisions (RHICs), quark stars etc., QGP 
consists of quarks and gluons confined in a volume large compared to 
the size of hadrons. In the case of early universe, just before 
hadronization, again we have QGP. 
In all above cases, EoS of QGP is needed to explain 
their evolution and study various properties. Since quantum chromodynamics 
(QCD) is nonlinear, 
it is difficult to solve analytically at low energy and only at high energy 
one can do perturbative calculations. That is why one goes to numerical 
simulations of QCD on lattice and EoS of QGP without and with quark flavors 
has been derived \cite{ka.1}. Many phenomenological models like 
hard thermal loop (HTL) \cite{reb.1}, quasi-particle QGP 
(qQGP) \cite{pe.1,ri.1}, strongly interacting QGP (sQGP) \cite{su.1}, 
strongly coupled QGP (SCQGP) \cite{ba.1} and various confinement models 
\cite{ba.2,ri.1,kd.1} were proposed to explain lattice 
results and there is no satisfactory model. Models which fits well with 
lattice results, generally, involve large number of parameters 
\cite{pe.1,ri.1} and 
models with detailed theory like HTL \cite{reb.1} don't fit 
lattice results well, 
so on. However, recently we found that SCQGP model \cite{ba.1} 
as well as a new qQGP model \cite{ba.3} fits surprisingly well 
with lattice results with minimum number 
of parameters compared to other models. Both models are purely 
phenomenological and based on the collective properties of plasma. 

Here we approach the problem from the known fact that RHO model of 
hadrons quite successfully explains many properties of hadrons \cite{vi.1}. 
QGP formed in RHICs and compact stars may be viewed as a large hadron with 
large number of quarks and gluons, confined in a RHO confining 
potential. Hence, a study of statistical mechanics and thermodynamics 
of such a system will give EoS of QGP. It is just a generalization 
of widely studied Bag Model of QGP. Instead of sharp bag wall, 
here we have 
a smooth confining potential. As a first step, here we neglect effects 
like Color-Coulomb interactions, hadron formations etc.  and see how the 
EoS of QGP differs from ideal EoS. 
Earlier, similar calculations 
has  been done \cite{kd.1}, but without proper counting of degeneracy. 
They have taken 
only accidental degeneracy where as here we have taken full degeneracy 
arising from coalese of continum levels to oscillator levels as a result 
of RHO interaction, just like in Landau's theory of diamagnetism \cite{pa.1}.       
\section{EoS of QGP in RHO Model :} 

From the earlier study of hadrons in RHO model, 
we have the single particle energy levels \cite{vi.1},
\begin{equation}
\epsilon_n^2 = C_g^2 \, (2 n + 1) \, ,
\end{equation}
for gluons and 
\begin{equation}
\epsilon_n^2 = m_q^2 + C_q^2 \sqrt{\epsilon_n + m_q}\, (2 n + 1) 
\, , \label{eq:en}  
\end{equation}
for quarks. $C_g$ and $C_q$ are spring constants of gluon and quark 
RHO confining potentials respectively. $n = 1,2,...$ is the 
oscillator quantum numbers and $m_q$ is the mass of quark.    
Let us first consider gluon plasma without quarks which is a Bose system. 
Following the standard procedure of statistical mechanics, pressure 
or logarithm of partition function is given by,
\begin{equation}
\frac{P\,V}{T} = - \sum_n g_n \, g_I \,\ln (1 -  e^{- \beta \epsilon_n}) \, ,  
\label{eq:p}
\end{equation}
where $g_n$ is the full degeneracy arising from the "coalescing together" 
of an almost continuous set of the zero-potential levels 
and is given by,
\begin{equation}
g_n = \frac{V C_g^3}{6 \pi^2} \, ( (2 n + 3)^{3/2} - (2 n + 1)^{3/2} ) 
\,. \label{eq:gn}   
\end{equation}
$V$ is the volume and $\beta$ is the inverse temperature. $g_I = 16$ is the 
internal degrees of freedom of gluons. Since gluons are bosons, 
the fugacity $z = 1$. For fermions like quarks $z \ne 1$ and pressure is 
given by,
\begin{equation}
\frac{P\,V}{T} = \sum_n g_n \, g_f \,\ln (1 + z\, e^{- \beta \epsilon_n}) \, ,  
\label{eq:pf}
\end{equation}
where $g_f$ is the internal degrees of quarks. 
Above expressions for pressure are exact 
and one need to do infinite sum which may be 
done numerically. However, for high temperature, we may use Euler-Maclaurin 
series to get an appproximate expression as, after some algebra,  
\[
\frac{P}{T} \approx \frac{g_I}{\pi^2}\,\sqrt{\frac{2}{\pi}}\, (C_g T)^{3/2} 
 \sum_{l=1}^{\infty} \frac{1}{l^{5/2}}\,\left[ K_{3/2} (\frac{C_g l}{T}) 
  + \frac{3}{4} \, \frac{C_g l}{T}\, K_{\frac{1}{2}} (\frac{C_g l}{T}) \right] 
\]
\begin{equation}
   + \frac{C_g^3 g_I}{12 \pi^2} \,(3 \sqrt{3} -1) \ln (1 - e^{- \frac{C_g}{T}}) 
\end{equation} 
for gluon plasma, in terms of modified Bessel functions $K_\nu (x)$. 
But we know that at extremely 
high temperature, confinement effects may be negligible and we expect to 
get EoS of relativistic free gas which we indeed get as shown below. 
As $T \rightarrow \infty$, 
\begin{equation}
P \rightarrow \frac{g_I \pi^2}{90}\,T^4 \,(1 + 
        \frac{45}{4\,\pi^2}\,\frac{C_g^2}{T^2} + ...) \, \label{eq:ps}
\end{equation}
    \[  \rightarrow  \frac{g_I \pi^2}{90}\,T^4 \equiv P_s \, .\]
Note that in our quantum calculation with proper choice of $g_n$, 
Eq. (\ref{eq:gn}), as expected, we get $T^4$, rather 
than $T^7$ as in \cite{kd.1}. Further, 
we also know that as $T \rightarrow \infty$, 
we approach classical limit and from our 
earlier classical calculations of this problem, we indeed got $T^4$ 
law \cite{ud.1}. 
Next let us compare our EoS with the results of lattice simulation of 
QGP or lattice gauge theory (LGT).   
\section{Comparison with LGT:} 

We can see, from Eq. (\ref{eq:ps}), that our EoS of QGP is 
temperature dependent modification of 
normal $T^4$ law. Similar modifications is seen in LGT calculations and 
only at $T \rightarrow \infty$, one expects $T^4$ law, 
which is termed as non-ideal 
effect. As an example, let us consider pure gauge QGP or gluon plasma and 
the pressure is, Eq. (\ref{eq:p}), 
\begin{equation}
\frac{P}{T} = - \frac{C_g^3 g_I}{6 \pi^2} \,
\sum_n ( (2 n + 3)^{3/2} - (2 n + 1)^{3/2} ) 
\,\ln (1 -  e^{- \frac{C_g}{T} \sqrt{2 n + 1}}) \, ,  \label{eq:p2}
\end{equation}
which may written as 
\begin{equation}
p \equiv \frac{P}{P_s} = - \frac{a^3\, g_I}{6 \pi^2} \, \frac{1}{t^3}\,
\sum_n ( (2 n + 3)^{3/2} - (2 n + 1)^{3/2} ) 
\,\ln (1 -  e^{- \frac{a}{t}\, \sqrt{2 n + 1}}) \, ,  \label{eq:p3}
\end{equation}
where $P_s$ is the Stefan-Boltzman limit pressure and $t \equiv T/T_c$. $T_c$ 
is the critical temperature of the transition of hadrons to QGP and 
the parameter $a$ is defined as, $a \equiv C_g / T_c$. Above 
equation may be compared with LGT results by varying the 
parameter $a$ as discussed in section V. Knowing $P$ or partition function, 
all other thermodynamic quantities like energy density, entropy, 
sound speed etc. may be derived. Further extension of our calculations 
for QGP with finite number of quark flavors, without or with finite 
chemical potential, is straight forward \cite{ud.2}.  

\section{Quark star:} 

As another example, let us consider what are called 
quark stars \cite{qs.1}. We know 
that, at the last stages of a star, it undergoes 
various changes like white-dwarf star, neutron star and then, 
probably, quark star. Sometimes 
more complex compact star with different layers of matter with quark matter 
inside and then mixture of quarks and hadrons, neutrons, so on. Here 
we take a very simple compact star, just cold quark star, where the 
degenerate pressure of quarks in RHO model balances the gravitational 
collapse. We estimate the limiting mass for stability 
and mass-radius relation, using Newtonian description. 
A detailed study using Tolman-Oppenheimer-Volkoff (TOV) equation 
with complex compact star in our model is a separate topic. With these 
simplifications, we get the degenerate pressure from the energy of the 
system,
\begin{equation}
E_0  = \frac{C_q^4\, V\, g_f}{6 \pi^2} \,
\sum_{n=1}^{n_F} ( (2 n + 3)^{3/2} - (2 n + 1)^{3/2} ) 
\,\sqrt{\bar{m}_q^2 + 2 n + 1} \, ,  \label{eq:p4}
\end{equation}
where we have approximated the single particle energy level of quarks, 
Eq. (\ref{eq:en}), for simplicity, as \cite{kd.1}  
\begin{equation}
\epsilon_n^2 = m_q^2 + C_q^2 \, (2 n + 1) \, .
\end{equation}
Here $g_f$ is the internal degrees of quarks 
and $\bar{m}_q \equiv \frac{m_q}{C_q}$.  
$n_F$ is the quantum number corresponding to Fermi level, given by the 
relation, 
\begin{equation}
N  = \frac{C_q^3\, V\, g_f}{6 \pi^2} \,
\sum_{n=1}^{n_F} ( (2 n + 3)^{3/2} - (2 n + 1)^{3/2} ) \, ,  
\end{equation}
where $N$ is the total number of quarks and in the case of cold star 
approximation, quarks fill the single particle RHO levels, according to 
Pauli's exclusion principle, upto the level $n_F$. Following the same 
procedure as that of white-dwarf star calculations we may relate the 
$n_F$ to the radius, $R$, of the star as 
\begin{equation}
R = \left[ \frac{9\,\pi\,M}{2\,m_q\,g_f\,C_q^3\, ( (2 n_F + 3)^{3/2} 
- 3 \sqrt{3}) } \right] ^{1/3}\, , 
\end{equation}
where $M$ is the mass of star. 
Degenerate pressure is obtained from $E_0$, 
\[ P = - \frac{\partial E_0}{\partial V} \] 
\begin{equation}
= - \frac{C_q^4\,  g_f}{6 \pi^2} \,
\left[ 1 - \frac{(2 n_F + 3)^{3/2} - 3 \sqrt{3} }{3 (2 n_F +3)^{1/2}} \, 
\frac{\partial }{\partial n_F} \right] 
\sum_{n=1}^{n_F} ( (2 n + 3)^{3/2} - (2 n + 1)^{3/2} ) 
\,\sqrt{\bar{m}_q^2 + 2 n + 1} \,. \label{eq:dp} 
\end{equation}
To proceed further, we approximate the summation over $n$ by integration 
using Euler-Maclaurin series. The error introduced by 
this approximation is small,  
which may be checked numerically by evaluating summation and Euler-Maclaurin 
formula. This degenerate pressure balances the gravitational pressure 
inward, which we take, the Newtonian formula, 
\begin{equation}
\frac{\alpha \, \gamma\,M^2}{4\,\pi\,R^4} \,,
\end{equation}     
where $\gamma$ is the gravitational constant and $\alpha$ is a geometrical 
factor.  After some lengthy calculations for large $n_F$, we get 
\begin{equation}
( \tilde{R}^2 + 1) (2 - \tilde{R}^2)^2 = \tilde{M}^{4/3} \,, \label{eq:qs} 
\end{equation} 
where $\tilde{R}$ and $\tilde{M}$ are related to radius, $R$, and mass, $M$, 
of the star respectively as, 
\begin{equation}
\tilde{R} = R\,C_q\,(\frac{2\,g_f\,m_q}{9\,\pi\,M})^{1/3} \,,
\end{equation} 
and     
\begin{equation}
\tilde{M} = \frac{M}{m_q}\, (\alpha\,\gamma\,m_q^2)^{3/2}\, 
\sqrt{\frac{g_f}{3\,\pi}}\,\,\frac{32}{9} \,.
\end{equation} 
It is interesting to compare above mass-radius relation with familiar 
white-dwarf star mass-radius relation, 
proceeding with similar normalization of mass and radius, 
\begin{equation}
( \tilde{R}^2 + 1) (2/3 - \tilde{R}^2)^2 = \tilde{M}^{4/3} \,. 
\end{equation} 
We see that the only change in our case is that the factor $2/3$ is 
replaced by $2$ in mass-radius relation and $C_q$ plays the role of 
$m_e$, the electron mass. From the Eq. (\ref{eq:qs}), we can estimate 
the limiting mass, $M_0^{QS}$, similar to the Chandrasekhar's limit ($M_0^W$)  
in the case of white-dwarf star, above which the star will collapse and we get,
\begin{equation}
M_0^{QS} = (\frac{m_p}{m_q})^2 \, \sqrt{\frac{32}{g_f}} \, M_0^W \, , 
\label{eq:mqs}   
\end{equation} 
where $m_p$ is the mass of proton. 
It is generally assumed \cite{qs.1} that quark star contains up, down and 
strange-antistrange quarks and hence $g_f \approx (2 \times 3 \times 2 
+ 1 \times 2\times 3 \times 2) = 24$. 
The quark mass $m_q$ in RHO is a constituent mass which ranges from 
$300$ $MeV$ to $600$ $MeV$ for up quark to strange quark and on an 
average we may approximate $m_q \approx m_p/2$. 
Therefore, as a rough estimate, from Eq. (\ref{eq:mqs}), 
limiting mass of quark star may be of the order of $4$ times the solar mass, 
where as the $M_O^W$  is of the order of solar mass. Further numerical 
calculation of Eq. (\ref{eq:qs}), without the approximation of 
$n_F \rightarrow  \infty$ which is used to get the Eq. (\ref{eq:qs}), 
we obtained a plot of mass-radius relation as shown Fig. 2. 

\section{Results :} 

In Fig. 1, we plotted our results for pressure in terms of $p \equiv P/P_s$ as 
function of $t \equiv T/T_c$ along with LGT results. By varying the 
parameter $a$, so as to get a similar behaviour as LGT results as 
function of $t$, we get $a=2.3$. However, the curve lies slightly 
above the lattice data. We may  correct our pressure by subtracting 
a constant $\Delta$ such that both the curve matches at large $t$ 
values and we get good fit except near to $T=T_c$. The value of $\Delta$ here 
is $0.1$. It is not unreasonable 
to subtract by $\Delta$ because we have neglected the mutual interactions 
between particles, mainly, attractive Color-Coulombic interactions. 
We know that attractive mutual interactions among particles reduces 
the pressure. From the  expression for $a$, we may estimate 
$T_c = C_g /a \approx 200\,MeV$ for $C_g = 2.35\,fm^{-1}$ \cite{kd.1}. 
Once pressure or partition function is obtained all other  
thermodynamic functions like $\varepsilon$, $c_s^2$ etc. may be 
easily evaluated. Similar comparison of our results and LGT results may be 
carried for other systems with flavored QGP \cite{ud.2}.  

In Fig. 2, we plotted mass-radius relations of quark star. Mass is expressed 
in terms of solar mass and radius in kilometers (km) by solving, numerically, 
Eq. (\ref{eq:dp}) after equating to the gravitational pressure. We have 
taken $\bar{m}_q = 2.5$ which is equivalent to taking 
$C_q \approx 1\,fm^{-1}$ \cite{kd.1}. 
  
\section{Conclusions :}

We developed statistical mechanics and thermodynamics of a system of 
quantum particles in RHO confining potential and applied to QGP. We 
compared the results with LGT and conclude that the overall QCD confining 
effect, modelled by RHO, explains the LGT data with small  
correction to take into account of the attractive mutual 
interactions among particles. We applied our model to study quark star 
and obtained the limiting mass and mass-radius relation and found to 
agree of the order of magnitude. To get an accurate result one need 
to study complex compact stars using TOV equations. However, Boson or 
Fermion system in RHO confining potential is a new general problem 
which may be applicable to many fields like BEC, RHICs, astrophysics 
and so on.     


\newpage
\begin{figure}
\caption { Plots of $P/P_s $ as a function of $T/T_c$ from 
our model and lattice results (symbols) for pure guage. } 
\label{fig 1}
\vspace{.75cm}

\caption { Mass-radius plot of quark star, mass in terms of solar mass and 
radius in kilometers.} 
\label{fig 2}
\vspace{.75cm}
\end{figure}

\begin{thebibliography}{99} 
\bibitem{qgp.1} Proceedings of 5$^{th}$ International Conference on 
Physics and AstroPhysics of Quark Gluon Plasma, Kolkata, India (2005), 
www.veccal.ernet.in/~icpaqgp. 
\bibitem{ka.1} G. Boyd, J. Engels, F. Karsch, E. Laermann, C. Legeland, 
M. Lutgemeier and B. Petersson, Phys. Rev. Lett. {\bf 75}, 4169 (1995);  
Nucl. Phys. {\bf B469}, 419 (1996); F. Karsch, Nucl. Phys. A, {\bf 698}, 
199 (2002); E. Laermann and O. Philipsen, Ann. Rev. Nucl. Part. Sci. {\bf 53}, 
163 (2003). 
\bibitem{reb.1} A. Rebhan, hep-ph/0010252, 2000.
\bibitem{pe.1} A. Peshier, B. Kampfer, O. P. Pavlenko and 
G. Soff, Phys. Lett. {\bf B337}, 235 (1994); Phys. Rev. {\bf D54}, 2399 (1996);
P.Levai and U. Heinz, Phys. Rev. {\bf C57}, 1879 (1998); R. A. Schneider 
and W. Weise, Phys. Rev. {\bf C64}, 055201 (2001). 
\bibitem{ri.1} D. H. Rischke, nucl-th/0305030, 2003.
\bibitem{su.1} E. Shuryak, hep-ph/0405066, (2004). 
\bibitem{ba.1} V. M. Bannur, Eur. Phys. J. {\bf C11}, 169(1999);hep-ph/0504072. 
\bibitem{ba.2} V. M. Bannur, Phys. Lett. {\bf B362}, 7 (1995). 
\bibitem{kd.1} S. B. Khadkikar, J. C. Parikh and P. C. Vinodkumar, 
Mod. Phys. Lett. {\bf A8}, 749 (1993). 
\bibitem{ba.3} V. M. Bannur, hep-ph/0508069. 
\bibitem{vi.1} S. B. Khadkikar and S. K. Gupta, Phys. Lett. {\bf B124}, 
523 (1983); Phys. Rev. {\bf D36}, 307 (1987); P. C. Vinodkumar, J. N. Pandya, 
V. M. Bannur and S. B. Khadkikar, Eur. Phys. J. {\bf A4}, 83 (1999).   
\bibitem{pa.1} R. K. Patria, {\it Statistical Mechanics}, 
ButterworthHeinemann, Oxford (1997). 
\bibitem{ud.1} K. M. Udayanandan, V. M. Bannur and P. C. Vinodkumar, 
submitted to J. Phys. G. 
\bibitem{ud.2} K. M. Udayanandan and V. M. Bannur, in preparation. 
\bibitem{qs.1} N. K. Glendenning, {\it Compact stars}, Springer, Berlin (2000);
 R. X. Xu, astro-ph/0211348; S. Banerjee, S. K. Ghosh and S. Raha, J. 
Phys. G: Nucl. Part. Phys., {\bf 26}, L1 (2000).   
\end{thebibliography}
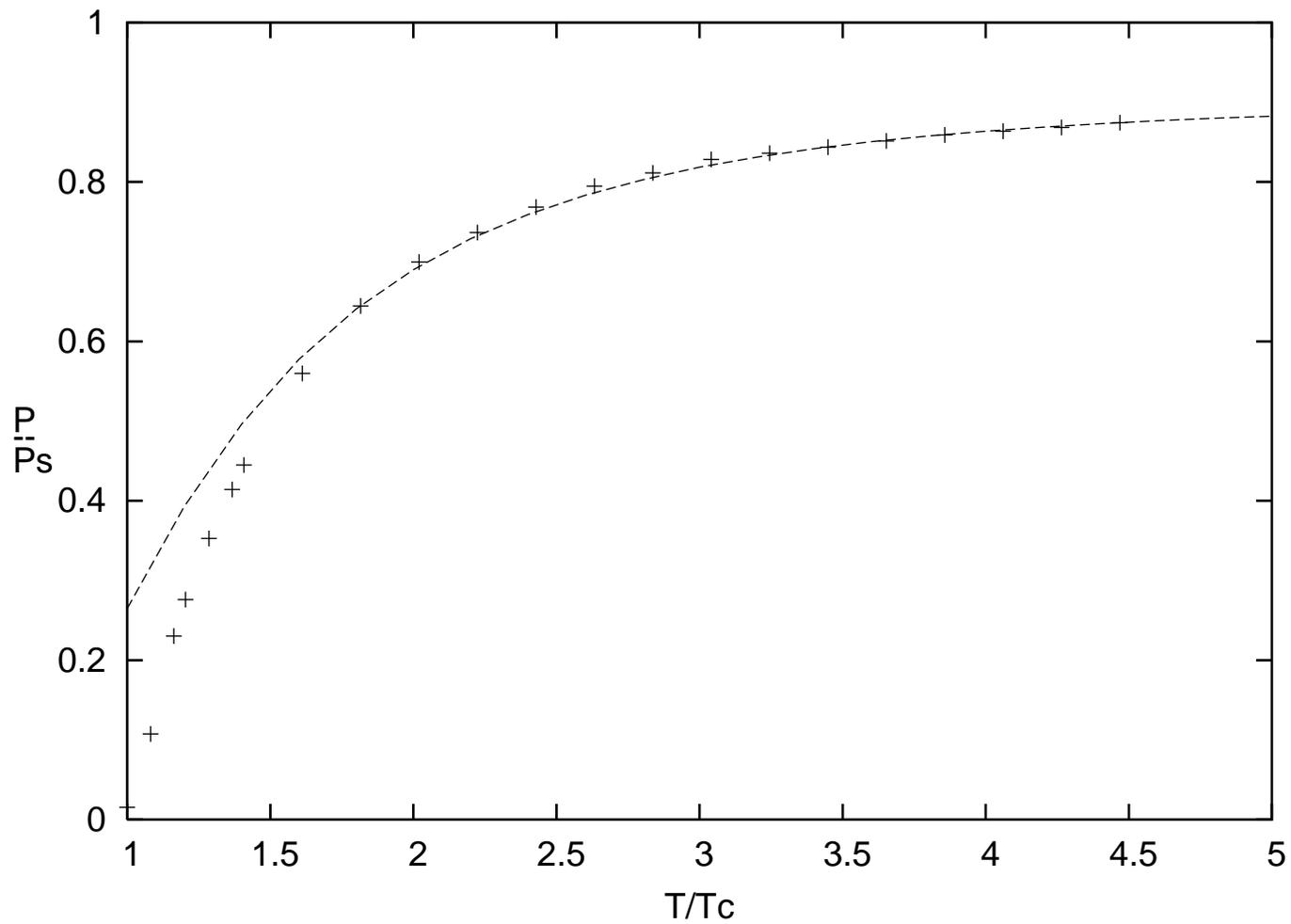
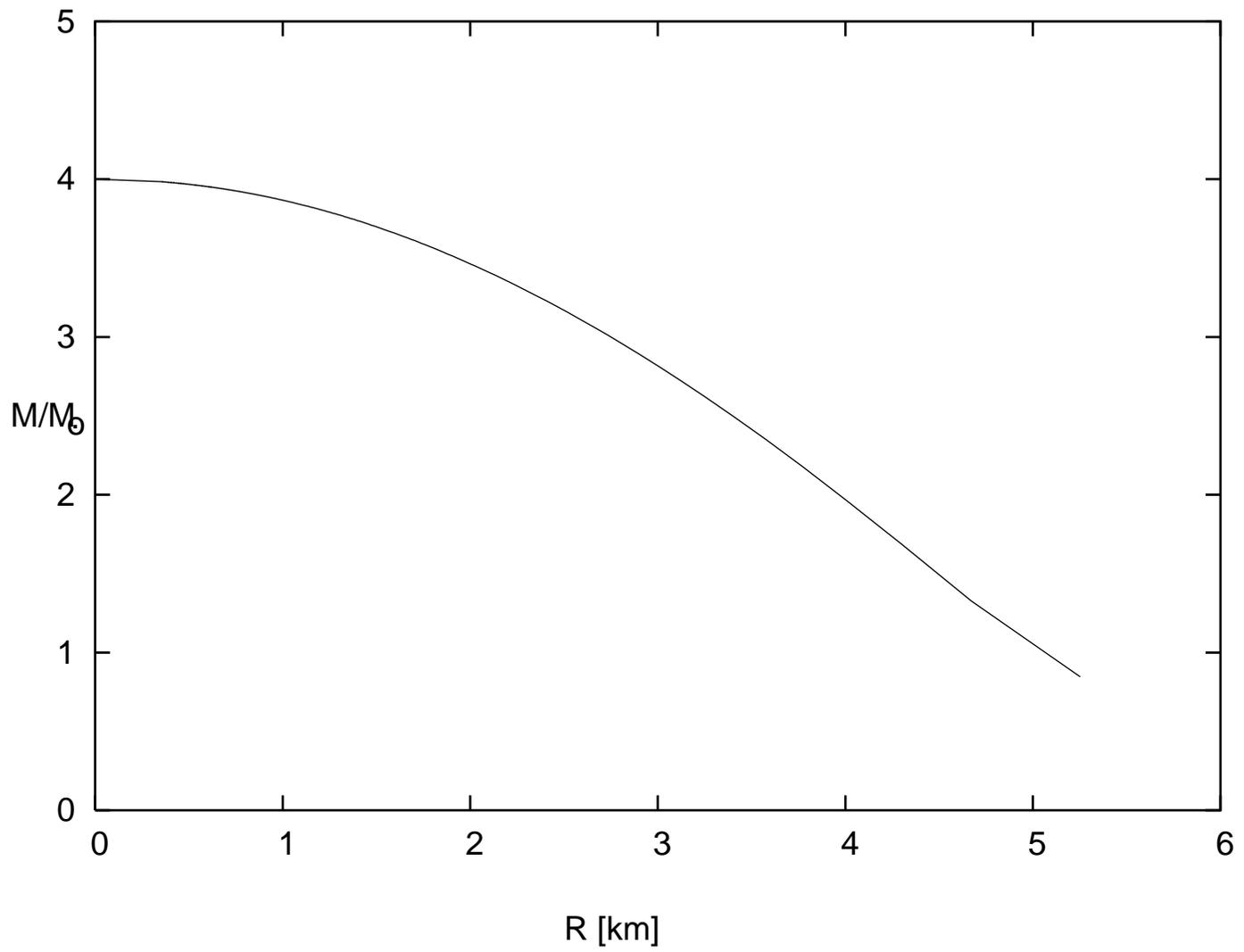
\end{document}